\documentclass[12pt]{iopart}
\usepackage{bm}
\usepackage{amsmath}
\usepackage{epsfig}
\usepackage{rotating}
\usepackage{graphicx}
\usepackage{ragged2e}
\usepackage{url}
\usepackage{caption}
\def\pr{\prime}
\def\be{\begin{equation}}

\def\ee{\end{equation}}
\def\barr{\begin{array}}
\def\earr{\end{array}}

\def\l{\left}
\def\r{\right}
\def\dis{\displaystyle}

\def\cc{{\cal C}}

\def\can{{\cal N}}

\def\cad{{\cal D}}

\def\la{\lambda}
\begin{document}

\title[$SU(3)$ irreps  for nucleons in a $\eta$-shell and nuclei with proxy-$SU(4)$ symmetry]{Irreducible representations of $SU(3)$ for nucleons in a oscillator shell $\eta$ with $\eta \le 8$: Application to nuclei with proxy-$SU(4)$ symmetry}

\author{V.K.B. Kota}
\address{Physical Research  Laboratory, Ahmedabad 380 009, India}
\ead{vkbkota@prl.res.in}
\vspace{10pt}
\begin{indented}
\item[]October 2026
\end{indented}

\begin{abstract}

Applying some simple rules, established in group theory literature, generated are codes (easy to modify as required) for obtaining $SU(3)$ irreducible representations (irreps) for a given number of nucleons in a given oscillator shell $\eta$ with $\eta \le 8$. In this approach, first the $SU(3)$ irreps are obtained for spinless fermions in a given $\eta$ shell. Using the codes developed, tables of $SU(3)$ irreps are generated for identical nucleons (protons or neutrons) with a given total spin $S$. Similarly, the $SU(3)$ irreps are generated for nucleons with a given spin-isospin $SU(4)$ irrep. Some applications of these tables for $\eta=4$ shell for $52 \le \mbox{Z,N} \le 78$ nuclei with proxy-$SU(4)$ symmetry are presented. Also, the procedure used for the shell model $SU(3)$ limits is easy to be extended for the $SU(3)$ limit of various interacting boson models and some examples for these are also presented.

\end{abstract}

\vspace{10pt}

%

\section{Introduction}

Elliott showed for the first time in 1958 that the oscillator orbital $SU(3)$ symmetry generates rotational spectra within the spherical shell model of atomic nuclei \cite{Ell-58a,Ell-58b}. Implicit here also is the goodness of Wigner's spin-isospin $SU(4)$ symmetry \cite{Wig-37,JCP,jpd-su4a,jpd-su4b}. Overcoming the breaking of $SU(3)$ symmetry due to the strong shell model spin-orbit force, $SU(3)$ model appeared in many different forms in nuclear structure. Some of these are: pseudo-$SU(3)$ model based on pseudo spin and pseudo Nilsson orbits, the proxy-$SU(3)$ model introduced by Bonatsos et al, $SU(3)$ in the $Sp(6,R)$ model, $SU(3)$ in the interacting boson, boson-fermion and boson-fermion-fermion models, $SU(3)$ in various cluster models and so on. For more details and references up to 2020 see \cite{su3-book}. See also \cite{su-1,su-2,su-3,su-4,su-5,su-6a,su-6b} for some of the recent applications of $SU(3)$ symmetry in nuclei. The focus of the present article is on the proxy-$SU(3)$ model and the related proxy-$SU(4)$ symmetry in nuclei.

Proxy-$SU(3)$ model was introduced in 2017 for heavy deformed nuclei with protons and neutrons occupying different shells \cite{Bona-1}. This scheme is defined through the replacement of intruder orbitals in a given shell by orbitals dropped into the lower shell of the same type of nucleons (protons or neutrons). Then, the intruder high-$(\eta, \ell, j)$ orbit that is pushed down due to the strong spin-orbit force is replaced by the proxy $(\eta-1,\ell-1,j-1)$ orbit by ignoring the high-lying $(\eta,\ell,j:k=\pm j)$ states. Then, for example $^0h_{11/2}$ in 50-82 shell changes to proxy $^0g_{9/2}$ with $k=\pm 11/2$ states of $h_{11/2}$ ignored. Similarly, the 50-82 shell becomes proxy $\eta=4$ shell, the 82-126 shell becomes proxy $\eta=5$ shell and so on. With the proxy oscillator shells, we have proxy-$SU(3)$ symmetry in each shell and by coupling these proton $SU(3)$ and neutron $SU(3)$ algebras, we have proxy-$SU(3)$ symmetry for heavy nuclei.
The proxy-$SU(3)$ scheme is quite successful in describing in a simple manner prolate shape dominance over oblate shape in nuclei \cite{Bona-2,Bona-3}. In addition, the proxy-$SU(3)$ model is applied in the study of preponderance of triaxial shapes in nuclei \cite{Bona-4,Bona-5}. Crucial element in these analysis is the calculation of the deformation parameters $(\beta , \gamma)$
that correspond to a given $SU(3)$ irreducible representation (irrep).
In addition to all these, more recently in \cite{KS-psu4} it was pointed out that proxy-$SU(4)$ symmetry is important in nuclei where the valence protons and neutrons occupy the same proxy oscillator shell. For these nuclei, goodness of proxy-$SU(3)$ model implies that proxy-$SU(4)$ is good.

In the applications of various $SU(3)$ models and in applying the proxy-$SU(3)$ model in particular, the first important problem to be solved is enumeration of $SU(3)$ irreps for a given number of nucleons in a given oscillator shell with major oscillator quantum number $\eta$ (in nuclear structure applications $\eta \le 8$). Many methods for obtaining the $SU(3)$ irreps are available in literature \cite{su3-book,Wy-70,Little,Pat-sh}. Also, computer codes for generating these are available \cite{unsu-1,unsu-2,unsu-3}. However, to have
a simple method that can be easily converted into a computer code
is certainly useful. This will also provide checks for the results generated using the codes in \cite{unsu-1,unsu-2,unsu-3}. One of the purposes of the present paper is to describe such a simple method. Using this, as presented in Section 2, a large set of tabulations of $SU(3)$ irreps are generated and these are also described in Section 2.  Secondly, using the $SU(3)$ reduction tables, as an application,
we consider nuclei with both valence protons and neutrons in the same shell, i.e. for nuclei with good proxy-$SU(4)$ symmetry. Results for nuclei with $32 \le \mbox{Z,N} \le 46$ are reported recently in \cite{Ko-nhw}. In Section 3, presented are the results for nuclei with $52 \le \mbox{Z,N} \le 78$ and also list the important questions that arise from these results. Answering these need further work in future. In Section 4, we will present briefly some results for the $SU(3)$ limits of various interacting boson models. Finally, Section 5 gives conclusions and future outlook. 

\section{Simple method for $SU(3)$ reductions in shell model}

Given $m$ number of nucleons in an oscillator shell $\eta$ with the nucleons carrying spin($S$) and isospin($T$) degrees of freedom, the spectrum generating algebra (SGA) is $U(4\can)$ where $\can=(\eta+1)(\eta+2)/2$. Then, for $\eta=2-8$ we have $\can$ taking values $6$, $10$, $15$, $21$, $28$, $36$ and $45$ respectively. Now, $SU(3)$ appears as a subalgebra in (called I)
$$
U(4\can) \supset \l[U(\can) \supset SU(3) \r] \otimes SU_{ST}(4)\;.
$$
The chains $SU(3) \supset SO_L(3)$ and $SU_{ST}(4) \supset SU_S(2) \otimes SU_T(2)$ are not discussed in this paper except when required. For identical nucleons ($m$ number of protons or neutrons), with only spin degree of freedom, similarly we have (called II),
$$
U(2\can) \supset \l[U(\can) \supset SU(3) \r] \otimes SU_{S}(2)\;.
$$
Finally, for spinless fermions (this system is mathematically
important as we will see ahead) we have (called III)
$$
U(\can) \supset SU(3)\;.
$$
In the Young tableaux notation, the irreps $\{f\}$ of $U(\can)$ for chains I, II and III will be of maximum four columns, two columns and one column respectively, i.e. they are of the type
$\{4^a3^b2^c1^d\}$, $\{2^a1^b\}$ and $\{1^a\}$ respectively. Similarly, the $SU(3)$ irreps are labeled by two numbers $(\la,\mu)$ in Elliott's notation. Our interest in this Section is to
obtain reductions of the irreps $\{f\}$ to $(\la,\mu)$ irreps. Towards this end, we use the following steps. 

\subsection{Kronecker products of $SU(3)$ irreps}

Firstly we need a simple method for evaluating the product (outer product or Kronecker product) of any two given $SU(3)$ irreps, i.e. determine
\be
(\la_1,\mu_1) \times (\la_2,\mu_2) = \dis\sum_{\la_3,\mu_3} \Gamma_{(\la_3,\mu_3)} \, (\la_3,\mu_3)\;.
\label{eq1}
\ee
Here, $\Gamma_{(\la_3,\mu_3)}$ gives the number of times the irrep
$(\la_3,\mu_3)$ appears in the Kronecker product. Note that in general $\Gamma_{(\la_3,\mu_3)} > 1$ giving the multiplicity of the irrep $(\la_3,\mu_3)$. Also, $\Gamma_{(\la_3,\mu_3)} = 0$ implies that the irrep $(\la_3,\mu_3)$ is absent. For solving Eq. (\ref{eq1}), first we need to find $(\la_3,\mu_3)$ that are all in principle allowed. For this, the equivalence is used between $U(3)$ irreps $\{g_1,g_2,g_3\}$ and the $SU(3)$ irreps $(\la,\mu)$. Note that given $\{g_1,g_2,g_3\}$, we have $\la=g_1-g_2$ and $\mu=g_2-g_3$. Also, $g_i$ are positive integers and $\{g_1,g_2,g_3\}$ are 3-row partitions of the integer $g$ with $g=g_1+g_2+g_3$ and $g_1 \ge g_2 \ge g_3 \ge 0$. Therefore, we can think of $(\la_1,\mu_1)$ as a $SU(3)$ irrep for $m_1=\la_1+2\mu_1$ number of particles and similarly $(\la_2,\mu_2)$ as a $SU(3)$ irrep for $m_2=\la_2+2\mu_2$ number of particles. Thus, $(\la_3,\mu_3)$ are all the $SU(3)$ irreps for $m_3=\la_1+2\mu_1+\la_2+2\mu_2$ number of particles. Then, we have easily
\be
\la_3=f_1-f_2\;\;,\mu_3=f_2-f_3\;\;\mbox{with}\;\;f_1+f_2+f_3=
\la_1+2\mu_1+\la_2+2\mu_2\;\;\mbox{and}\;\;f_1 \ge f_2 \ge f_3 \ge 0\;.
\label{eq2}
\ee
Having determined all the $(\la_3,\mu_3)$ irreps, it is easy to find the value of $\Gamma_{(\la_3,\mu_3)}$ in Eq. (\ref{eq1}) for any given $(\la_3,\mu_3)$ using the rules given by Chew and Sharp \cite{CS}. In fact these rules can be converted into a simple program and this is already available, as MULTU routine, in Akiyama and Draayer $SU(3)$ Wigner-Racah package \cite{jpdaky}. 
In summary, following the steps described above, it is easy to evaluate Kronecker product of any two $SU(3)$ irreps.

Before proceeding further, it is useful to mention that given $m$ particles in a $\eta$ shell, all the $SU(3)$ irreps $(\la,\mu)$ that are in principle possible for the $m$ particle system are given by
\be
\barr{l}
(\la,\mu)=(f_1-f_2,f_2-f_3)\;;\\
f_1+f_2+f_3=m\eta\;,\;\;\; f_1 \ge f_2 \ge f_3 \ge 0\;.
\earr \label{eq2a}
\ee 
Note that $\{f_1,f_2,f_3\}$ are $U(3)$ irreps in Young tableaux notation.

\subsection{Reductions of totally antisymmetric irreps of $U(\can)$ to $SU(3)$ irreps}

In the second step all the antisymmetric irreps $\{1^m\}$ of $U(\can)$ will be reduced to $SU(3)$ irreps and this corresponds to chain III above. Note that $m$ takes values
$0$ to $\can$; $\can = (\eta+1)(\eta+2)/2$. Using particle-hole
symmetry, we have
\be
\l\{1^r\r\} \rightarrow \dis\sum_i x_i (\la_i , \mu_i) \oplus 
\Rightarrow \l\{1^{\can -r}\r\} \rightarrow \dis\sum_i x_i (\mu_i , \la_i) \oplus\;.
\label{eq3}
\ee
Here, $x_i$ is the multiplicity of the irrep $(\la_i , \mu_i)$ implying that the irrep $(\la_i , \mu_i)$ appears $x_i$ times [$x_i=0$ implies that the irrep $(\la_i , \mu_i)$ will not appear in the reduction]. Thus, we need only the reductions for $\{1^m\}$ with $m \le [\can/2]$ where $[\can/2]$ is the integer part of $\can/2$. For these, the $x_i$ are determined using the following procedure. Given a $\eta$ oscillator shell, the single particle (sp) states in cartesian co-ordinates are denoted by $(n_z,n_x,n_y)$ with $n_i \ge 0$ and $n_z+n_x+n_y=0$. Note that $n_i$ gives the quanta in $i$-th direction. Also, $n_z=\eta-r$, $n_x=r-s$ and $n_y=s$ with $r=0$ to $\eta$ and $s$ is $0$ to $r$. We will arrange the $(n_z,n_x,n_y)$ states with increasing value of $r$ and for a given $r$ with increasing value of $s$. Then for example, for $\eta=3$ the 10 sp states are $(3,0,0)$, $(2,1,0)$,
$(2,0,1)$, $(1,2,0)$, $(1,1,1)$, $(1,0,2)$, $(0,3,0)$, $(0,2,1)$,
$(0,1,2)$ and $(0,0,3)$. Now, these states are filled in all possible ways with the $m$ particles such that there is a maximum of one particle in any state. Each distribution giving an $m$-particle configuration, we can calculate for each of them the quanta $N_z=\sum n_z$, $N_x=\sum n_x$ and $N_y=\sum n_y$ where the sum is over the occupied $(n_z,n_x,n_y)$ states. Collecting $(N_z,N_x,N_y)$ for all the $\binom{\can}{m}$ configurations, it
is easy to determine $d(N_z,N_x,N_y)$ giving the number of times a $(N_z,N_x,N_y)$ appears. Note that given $m$ and $\eta$, the
$(N_z,N_x,N_y)$ follow from the fact that $N_r$'s with $r=z,x$ and $y$ are positive integers and $N_z+N_x+N_y=m\eta$. Having determined all $d(N_z,N_x,N_y)$, it is easy to obtain $D(\la^\prime , \mu^\prime)$ where $\la^\prime=N_z-N_x$ and $\mu^\prime = N_x-N_y$. Now, the number of times a given $SU(3)$ irrep $(\la,\mu)$ appears in $\{1^m\}$ of $U(\can)$ follows from the difference formula \cite{Bard-su4,Bloc-su4,Kene-66},
\be
\cad(\la,\mu)= D(\la,\mu) - D(\la-1,\mu+2) - D(\la+2,\mu-1) + D(\la+3,\mu) + D(\la,\mu+3) - D(\la+2,\mu+2) \;.
\label{eq4}
\ee                 
In order to implement the procedure described here on a computer, all we need is an efficient program for enumerating all the $(N_z,N_x,N_y)$ configurations that are $\binom{\can}{m}$ in number. Then, using Eq. (\ref{eq4}) gives easily $\cad(\la,\mu)$ and this is the same as $x_i$ in Eq. (\ref{eq3}). We have developed a simple code for implementing this procedure and obtained complete set of reductions for $\eta=3$ to $7$ and these can be extended to $\eta=8$ when needed (at present we have $\eta=8$ results for $m \le 10$). In \cite{su3-book}, complete tables for $\eta=3$, $4$ and $5$ are presented. It is easy to obtain the results for $\eta=2$ and they are
\be
\{0\}_{\eta=2} = (0,0),\;\;\{1\}_{\eta=2} = (2,0),\;\;\{1^2\}_{\eta=2} = (2,1),\;\;\{1^3\}_{\eta=2} = (3,0 ) \oplus (0,3)\;.
\label{eq5}
\ee
For $\{1^m\}$ with $m=4$, $5$ and $6$ the reductions follow from Eq. (\ref{eq3}). Complete Tables for $\eta=6$ and $\eta=7$ and for $m \le 10$ for $\eta=8$ can be obtained from the author. 

\subsection{Reduction of two and four column irreps of $U(\can)$ to $SU(3)$ irreps}

\subsubsection{Reduction of two column irreps}

Nuclei with valence protons and valence neutrons occupying different oscillator shells $\eta_p$ and $\eta_n$ respectively,
we have chain II for protons with $\can \rightarrow \can_p = (\eta_p+1)(\eta_p+2)/2$ and similarly for neutrons with $\can \rightarrow \can_n = (\eta_n+1)(\eta_n+2)/2$. Given the number of valence protons $m_p$ and spin say $S_p$, the $U(\can_p)$ irrep $\{f_p\}$ is uniquely given by $\{f_p\} = \{2^a1^b\}$ with
$a=(m_p-2S_p)/2$ and $b=2S_p$. Then, we need $\{2^a1^b\}$ reduction to $SU(3)$ irreps $(\la_p,\mu_p)$, i.e.
\be
(m_p,S_p) \leftrightarrow \l\{2^a1^b\r\} = \dis\sum_{\la_p,\mu_p} \Gamma_{(\la_p,\mu_p)}\;(\la_p,\mu_p)
\label{eq6}
\ee
with $\Gamma_{(\la_p,\mu_p)}$ giving the multiplicity of the $SU(3)$ irrep $(\la_p,\mu_p)$. For obtaining the $\Gamma$'s, first we need to enumerate all $(\la_p,\mu_p)$. For a given $m_p$ value, they are all given by Eq. (\ref{eq2a}). Now, we will use the law \cite{Wy-70,Little}
\be
\l\{2^a1^b\r\} = \l\{1^{a+b}\r\} \times \l\{1^a\r\} - \l\{1^{a+b+1}\r\} \times \l\{1^{a-1}\r\} \;.
\label{eq7}
\ee
Therefore, we have terms of the type $\{1^{X_1}\} \times \{1^{X_2}\}$ and they have to be reduced to $SU(3)$ irreps. To this end we start with, obtained via step 2 described above,
\be
\barr{l}
\l\{1^{X_1}\r\} = \dis\sum_i x_i (\la_{1i} , \mu_{1i})\;, \\
\l\{1^{X_2}\r\} = \dis\sum_j y_j (\la_{2j} , \mu_{2j})\;. 
\earr \label{eq8}
\ee
Then we need, the following Kronecker product reduction,
\be
(\la_{1i} , \mu_{1i}) \times (\la_{2j} , \mu_{2j}) \rightarrow
\dis\sum_k \kappa^{(ij)}_{12:k} (\la_{12:k} , \mu_{12:k})\;.
\label{eq9}
\ee
The $(\la_{12:k} , \mu_{12:k})$ here are all the $SU((3)$ irreps for $X_{12}=X_1+X_2$ number of particles and they are enumerated using Eq. (\ref{eq2a}). The multiplicity coefficients $\kappa^{(ij)}_{12:k}$ follow easily from the rule given by Chew and Sharp \cite{CS}; see step 1 above. Then we have,
\be
\barr{l}
\l\{1^{X_1}\r\} \l\{1^{X_2}\r\} = \dis\sum _k z_k (\la_{12:k} , \mu_{12:k})\;;\\
z_k = \dis\sum_{i,j} x_i y_j \kappa^{ij}_{12:k}\;.
\earr \label{eq10}
\ee
Applying Eq. (\ref{eq10}) to the two terms in Eq. (\ref{eq7}) and
simplifying will give Eq. (\ref{eq6}). Repeating the same procedure for $m_n$ number of neutrons in a $\eta_n$ shell with spin $S_n$ will give
\be
(m_n,S_n) \leftrightarrow \l\{2^c1^d\r\} = \dis\sum_{\la_n,\mu_n} \Gamma_{(\la_n,\mu_n)}\;(\la_n,\mu_n)
\label{eq11}
\ee
For the combined proton-neutron system, then we have $SU_p(3) \otimes SU_n(3) \supset SU(3)$ and the $SU(3)$ irreps with associated multiplicities are obtained using Eqs. (\ref{eq6}) and (\ref{eq11}) and by performing the Kronecker products of the proton and neutron $SU(3)$ irreps (here step 1 is needed).

\subsubsection{Reduction of four column irreps}  

Nuclei with valence protons and neutrons in the same oscillator shell $\eta$, the Wigner's spin-isospin $SU(4)$ symmetry is important, i.e. chain I applies. Now, given the nucleon number $m$ and the isospin $T=|T_Z|$ (note that $T_Z=(m_p-m_n)/2$ where $m_p$ is number of valence protons and $m_n$ is number of valence neutrons with $m=m_p+m_n$), the lowest $U(4)$ irrep $\{F_1,F_2,F_3,F_4\}$ is given as follows (note that $SU(4)$ irreps follow from $U(4)$ irreps). In this paper we will restrict to the situation with $m$ always even. Then we have \cite{JCP,MK-su4,Piet-su4},
\be
\barr{l}
\l\{F_1,F_2,F_3,F_4\r\} = \l\{\frac{m+2T}{4}, \frac{m+2T}{4}, \frac{m-2T}{4}, 
\frac{m-2T}{4}\r\}\;\;\;\mbox{for}\;\;\;\frac{m}{2}+T\;\;\mbox{ even}\;,\\
\l\{F_1,F_2,F_3,F_4\r\}=\l\{\frac{m+2T+2}{4}, \frac{m+2T-2}{4}, 
\frac{m-2T+2}{4}, \frac{m-2T-2}{4}\r\}\;\;\;\mbox{for}\;\;\;
\frac{m}{2}+T\;\;\mbox{odd}\;.\\
\earr \label{eq12}
\ee
The only exception is $T=0$ for $m=4r+2$ type and then 
\be
\l\{F_1,F_2,F_3,F_4
\r\}=\l\{\frac{m+2}{4}, \frac{m+2}{4}, \frac{m-2}{4}, \frac{m-2}{4}\r\}\;.
\label{eq13}
\ee
Using  Eqs. (\ref{eq12}) and (\ref{eq13}) it is easy to obtain the lowest $U(4)$ irrep for a given $m$  and $T=|T_z|$, i.e. for a given even-even nucleus. With the $U(4)$ irrep being four rowed, from I it is easy to see that $\{f\}_{U(\can)}$ will be four columned. Then, $\{f\}$ takes the form $\{4^a 3^b 2^c 1^d\}$ with $a=F_4$, $b=F_3-F_4$, $c=F_2-F_3$ and $d=F_1-F_2$. with the reductions for $\{1^r\}$ of $U(\can)$ are all available (see step 2 above), the reductions for any 4 column irrep $\{f\}$ of $U(\can)$ to $SU(3)$ irreps is obtained by first 
expanding $\{f\}$ as a $4 \times 4$ determinant involving only
antisymmetric irreps of $U(\can)$. Given $\{f\}=\{4^a 3^b 2^c 1^d\}$ irrep, we have \cite{Wy-70,Little},
\be
\barr{l}
\l\{4^a 3^b 2^c 1^d\r\} = \l|\barr{cccc} \l\{1^{F_1}\r\} &  
\l\{1^{F_1+1}\r\} &  \l\{1^{F_1+2}\r\} & \l\{1^{F_1+3}\r\} \\
\l\{1^{F_2-1}\r\} &  \l\{1^{F_2}\r\} &  \l\{1^{F_2+1}\r\} & \l\{1^{F_2+2}\r\} \\ \l\{1^{F_3-2}\r\} &  \l\{1^{F_3-1}\r\} &  \l\{1^{F_3}\r\} & \l\{1^{F_3+1}\r\} \\ \l\{1^{F_4-3}\r\} &  
\l\{1^{F_4-2}\r\} &  \l\{1^{F_4-1}\r\} & \l\{1^{F_4}\r\} \earr \r|\;;\\
F_1=a+b+c+d,\;\;F_2=a+b+c,\;\;F_3=a+b,\;\;F_4=a\;.
\earr \label{eq14}
\ee
The multiplications in the determinant in Eq. (\ref{eq14}) are Kronecker multiplications. Also, $\{1^x\}=0$ if $x < 0$ and $\{1^x\}=\{0\}$ if $x=0$. It important to note that when we expand the determinant in Eq. (\ref{eq14}), there will be 24 terms of the type 
$$
\l\{1^{X_1}\r\} \l\{1^{X_2}\r\} \l\{1^{X_3}\r\} \l\{1^{X_4}\r\}
$$
where $X_1+X_2+X_3+X_4=m$ if $\{f\}$ is an irrep of a $m$ nucleon system. First we will reduce $\{1^{X_1}\} \times \{1^{X_2}\}$ to $SU(3)$ irreps using Eq. (\ref{eq10}). Starting with Eq. (\ref{eq10}) and $\l\{1^{X_3}\r\}$ decomposition to $SU(3)$ irreps $(\la_{3\ell} , \mu_{3\ell})$
with multiplicities say $p_{3\ell}$, and repeating the procedure that gave Eq. (\ref{eq10}) but with the $SU(3)$ irreps $(\la_{123:l} , \mu_{123:l})$ for $X_{123}=X_1+X_2+X_3$ number of particles, we will obtain
$$
\l\{1^{X_1}\r\} \l\{1^{X_2}\r\} \l\{1^{X_3}\r\}
$$
reduction to $SU(3)$ irreps. Repeating the same procedure we will finally obtain the reduction of 
$$
\l\{1^{X_1}\r\} \l\{1^{X_2}\r\} \l\{1^{X_3}\r\} \l\{1^{X_4}\r\}
$$
to $SU(3)$ irreps with the associated multiplicities. Applying this to the 24 terms in the determinant in Eq. (\ref{eq14}) and adding them we will have the final result for a 4-column $\{f\}$ reduction to $SU(3)$ irreps. 

\subsection{Summary}

Following the steps described in Sections 2.1, 2.2 and 2.3, we have developed codes (in FORTRAN) for reducing any one, two and four columned irrep $\{f\}$ of $U(\can)$, with $\can=(\eta+1)(\eta+2)/2$, into $SU(3)$ irreps. As needed in many applications, in the current form the codes are used for irreps of the type $\{1^m\}$, $\{2^r\}$ and $\{4^r2^s\}$ with $\eta$ up to $7$ and in some situations for $\eta=8$. These codes are easy to change for larger $\eta$ values and for any specific type of $\{f\}$. These codes are available upon reasonable request.

Final results for the reduction of any $\l\{f\r\}$ of $U(\can)$ are of the form,
\be
\l\{f\r\} = \dis\sum_q \Gamma_q \l(\la_q , \mu_q\r) \oplus \;.
\label{eq15}
\ee
where $\Gamma_q$ is the multiplicity of the $SU(3)$ irrep $\l(\la_q , \mu_q\r)$ and the irreps $\l(\la_q , \mu_q\r)$ are
all the $SU(3)$ irreps for $m=\sum_{i=1}^{\can}\; f_i$. A good check of the reductions obtained is the dimensionality check.
A simple formula for $U(\can)$ irreps dimension and $SU(3)$ irreps dimension along with Eq. (\ref{eq15}) will give \cite{Wy-70,Little},
\be
d(\l\{f\r\}) = \dis\prod_{i<j=1}^{\can} \dis\frac{f_i-f_j+j-i}{j-i} = \dis\sum_q \Gamma_q \;(\la_q+1)(\mu_q+1)(\la_q +\mu_q+2)/2\;.
\label{eq15a}
\ee
This equation forms a good (and important) check for the reductions obtained using a computer code. Also, in applications
we need to arrange in the order of  decreasing weight the $SU(3)$ irreps that are obtained. Note that an irrep $(\la , \mu)$ is said to be of higher weight than the irrep $(\la^\pr , \mu^\pr)$ [then we say $(\la , \mu) > (\la^\pr , \mu^\pr)$] if either of the following is satisfied,
\be
\barr{l}
(i)\;\;(\la , \mu) > (\la^\pr , \mu^\pr)\;\;\mbox{if}\;\;2\la+\mu > 2\la^\pr +\mu^\pr\;,\\
\mbox{or} \\
(ii)\;\;(\la , \mu) > (\la^\pr , \mu^\pr)\;\;\mbox{if}\;\;2\la+\mu = 2\la^\pr +\mu^\pr\;\;\mbox{and}\;\;\mu > \mu^\pr \;.
\earr \label{eq16}
\ee 
In the set of $(\la_q,\mu_q)$, the highest weight (hw) irrep is denoted by $(\la_H , \mu_H)$. In the following we will also use $\epsilon_H = 2\la_H +\mu_H$. A simple formula for the hw irrep for any $\eta$ and any $\{f\}$ of $U(\can)$ [$\can = (\eta+1)(\eta+2)/2$] is as follows \cite{su3-book,Ko-hw}, 
\be 
\barr{l} \l\{f\r\}_{U((\eta
+1)(\eta +2)/2)} \rightarrow (\lambda_H , \mu_H)_{SU(3)}  \;\;\mbox{with} \\
\lambda_H = \dis\sum_{r=0}^\eta \dis\sum_{x=0}^r (\eta -2r+x) \times 
f_{1+x+\frac{r(r+1)}{2}}\;,\;\;\;\;\mu_H=\dis\sum_{r=0}^\eta \dis\sum_{x=0}^r 
(r-2x) \times f_{1+x+\frac{r(r+1)}{2}}\;. 
\earr 
\label{eq17} 
\ee 
The $(\la_H,\mu_H)$ irrep has many applications and in particular when one uses the proxy-$SU(3)$ model \cite{Bona-2,Bona-3,Bona-4,Bona-5}.

\section{Applications to nuclei with $52 \le \mbox{Z,N} \le 78$ and proxy-$SU(4)$ symmetry}

Tables giving reduction of four column irreps of $U(\can)$ to $SU(3)$ irreps are needed for nuclei with good proxy-$SU(4)$ symmetry. This is valid, with good proxy-$SU(3)$ symmetry, for nuclei with both valence protons and neutrons occupying the same proxy oscillator shell $\eta$. Recently, nuclei with valence protons an neutrons in proxy $\eta=3$ shell are considered in some detail \cite{Ko-nhw}. Here, we will consider isotopes of Te, Xe, Ba, Ce, Nd, Sm, Gd, Dy, Er, Yb, Hf, W, Os and Pt with $52 \le \mbox{N} \le 78$. In these nuclei, valence protons and neutrons
occupy the proxy $\eta=4$ shell. Then we have $\can=15$. With proxy $SU(3)-SU(4)$ symmetry we have the algebra
$$
U(60) \supset \l[SU_{p+n}(15) \supset SU_{p+n}(3) \supset SO_{p+n:L}(3) \r] \otimes \l[SU_{p+n}(4) \supset SU_{p+n:S}(2) \otimes SU_{p+n:T}(2)\r]\;.
$$
Starting with a given nucleus, we have the number $m=m_p+m_n$ of valence nucleons, isospin $T=|m_p-m_n|/2$ and the corresponding lowest proxy-$SU(4)$ irrep given by Eqs. (\ref{eq12}) and (\ref{eq13}); note that $S=0$. The proxy-$SU(4)$ irreps give the corresponding $U(15)$ irreps $\l\{f\r\} = \l\{4^a,2^b\r\}$. In the situation that $m > 30$, we employ the corresponding hole irrep $\l\{f_h\r\}$ and for $a \ne 0$,
\be
\l\{f\r\} = \l\{4^a, 2^b\r\} \Rightarrow \l\{f_h\r\} =
\l\{4^{15-a-b},2^b\r\}\;.
\label{eq18}
\ee 
With this we use
\be
\l\{f_h\r\} \rightarrow \dis\sum_i x_i (\la_i , \mu_i) \oplus 
\Rightarrow \l\{f\r\} \rightarrow \dis\sum_i x_i (\mu_i , \la_i) \oplus\;.
\label{eq19}
\ee
Using the method described in Section 2, $SU(3)$ irreps are obtained for Te, Xe, Ba, Ce, Nd, Sm, Gd, Dy, Er, Yb, Hf, W, Os and Pt isotopes with
$52 \le \mbox{N} \le 78$. These irreps are ordered with decreasing weight
and given in the Tables 1 to 14 are the $SU(3)$ irreps $(\la,\mu)$ with $2\la+\mu=\epsilon_H$, $\epsilon_H-3$ and $\epsilon_H-6$. Thus, the Tables 1 to 14 give not only the hw irrep, but also next highest weight (nhw) irrep, next to next highest weight (nnhw) irrep and so on. Full tables with all $(\la,\mu)$ can be obtained from the author.
Using these, the deformation parameter $\beta$ and triaxiality parameter $\gamma$ can be investigated for the even-even Te to Pt isotopes with $52 \le \mbox{N} \le 78$ and assuming some mixing between the $SU(3)$ irreps.  
Given a $SU(3)$ irrep $(\la , \mu)$, the corresponding Bohr-Mottelson quadrupole shape parameters $(\beta,\gamma)$ are given by the following formulas \cite{Bona-4,Bona-5,beta-gam},
\be
\barr{rcl}
\beta & = & \cc \;\l[\dis\frac{4\pi}{5\l(A\overline{r^2}\r)^2}\,\l(\la^2 + \mu^2 + 
\la \mu + 3\la + 3\mu +3\r)\r]^{1/2}\;,\\
\gamma & = & \arctan \l[\dis\frac{\dis\sqrt{3}(\mu +1)}{(2\la +\mu +3)}\r]\;.
\earr \label{eq20a}
\ee 
Here, $A$ is nucleon number and $\overline{r^2}$ is the dimensionless mean square radius. It is well known that $(\overline{r^2})^{1/2} = r_{r.m.s} = r_0 A^{1/6}$ where $r_0 = 0.87$. Similarly, in Eq. (\ref{eq20a}) the scale factor $\cc$ arises as we are using only the valence nucleons. As used in \cite{Bona-4,Bona-5}, we have $\cc=A/(S_p+S_n)$ where $S_p$ and $S_n$ are the sizes of the proton and neutron valence shells. For the Te to Pt isotopes used here, $\cc=A/64$ as $S_p=S_n=32$ (protons and neutrons are in 50-82 shell). It is important to note that non-zero value of the $\gamma$ parameter indicates triaxiality. Tables with $(\beta,\gamma)$ values using only
the hw irrep $(\la_H,\mu_H)$ and also nhw-ee irrep are listed in \cite{Bona-4} for Te to Pt isotopes.
The nhw-ee irrep is defined as the $(\la \mu)$ next lower to the hw irrep but with both $\la$ and $\mu$ even. From Tables 1-14 it is easy read off both the hw and nhw-ee irreps. Plots of $\beta$ vs N and $\gamma$ vs N for Te to Pt isotopes with $52 \le \mbox{N} \le 78$ are shown in Figs. 1-4. Some comments on the results in the tables and the figures are as follows.

\begin{enumerate}

\item From the Tables 1-14 it is seen that in many situations the
nhw irrep is with either $\la$ odd (called nhw-oe) or $\mu$ odd
(called nhw-eo) or in some cases both $\la$ and $\mu$ odd (called nhw-oo). These will give rotational bands with the $K$
quantum number odd and they will not generate L=0 (with $S=0$, then $J=0$) levels. Therefore they will not mix with the ground $K=0$ band (assuming $K$ a good quantum number). This follows from the Elliott's classification \cite{Ell-58a,Ell-58b},
\be
\barr{rcl}
(\lambda \mu) \longrightarrow L: & & \\
K & = & min(\lambda,\mu),\; min(\lambda,\mu) -2,\; \cdots , 0\;\mbox{or}\;1, \\
L &=& K,\; K+1, \;K+2,\; \cdots , \; K + max(\lambda,\mu)\;\; for\;\;
K \neq 0\;, \\
L & = & max(\lambda,\mu),\;max(\lambda,\mu)-2,\; \cdots , 0\;\mbox{or}\; 
1\;\;for\;\; K = 0\;.
\earr \label{eq20}
\ee
For nuclei with hw irrep of the type $(\la,0)$, the mixing with the nhw is in particular important in the study triaxial shapes.
Though in many recent studies the nhw-eo, nhw-oe and nhw-oo are ignored \cite{Bona,Bona-4,Bona-5}, it should be noted that these irreps generate $2^+$ states and they can mix with the $2^+$ state of the ground band. This is important as $2^+_1$ and $2^+_2$ states give information about triaxiality \cite{Bona-tri} and
therefore nhw-eo, nhw-oe and nhw-oo irreps may prove to be important.
 
\item The nhw as well as nhw-ee in general will have multiplicity, i.e. the same $SU(3)$ irrep will appear more than once. Then, to distinguish these, i.e. to resolve multiplicity, we need some additional operators that are different from the $SU(3) \supset SO(3)$ integrity basis (IB) operators. The IB operators are the quadratic and cubic Casimir operators of $SU(3)$ and a third order and fourth order operator that are used for example in \cite{Jpd-We} and also in deriving Eq. (\ref{eq20a}). It is possible to distinguish the different $SU(3)$ irreps with the same $(\la,\mu)$ values by constructing operators similar to the one constructed and applied by Akiyama in the $sdg$IBM \cite{Aki-85}. As we have fermions in shell model and following \cite{Aki-85}, a possible operator is
\be
W=\l\{A^\dagger(04) \l[A^\dagger(04)\r]^{\dagger}\r\}^{(00)}\;\;\mbox{with} \;\; A^\dagger(04) = \l(a^\dagger_{(40)} a^\dagger_{(40)}\r)^{(04)} 
\label{eq21}
\ee
where $a^\dagger$ is a nucleon creation operator. For brevity, the spin and isospin labels are not shown for the $A^\dagger(04)$
operator. It will be a $(S_0T_0)=(10) \oplus (01)$ tensor. Also, the $W$ operator will be a spin-isospin scalar. Usefulness of the operator $W$ remains to be tested. Also, it is important to identify other operators, with good physical meaning, that can split the $SU(3)$ irreps with same $(\la,\mu)$. These operators will be a combination of $SU(3)$ scalars in $U(15)$. 

\item In the Tables 1-14 used is the proxy-$SU(4)$ irrep as appropriate for low-lying levels and determined by the value of isospin $T=|T_z|=(m_p-m_n)/2$. Then, we can easily identify from the tables the nhw-ee. For example $^{108-130}$Xe they are
$(20,6)^2$, $(24,8)$, $(32,2)^2$, $(34,2)^2$, $(28,12)$, $(34,4)^2$, $(32,6)^3$, $(32,4)^2$, $(22,16)$, $(26,6)^2$, $(20,10)^3$ and $(16,10)^3$ respectively. Interestingly, these irreps (without giving the multiplicities), as well as the hw irrep that follows from Eq. (\ref{eq17}) (this gives the same results as listed in Tables 1-14), also follow from a prescription that Bonatsos et al employed for determining the hw and nhw-ee for nuclei with valence protons and valence neutrons in different
oscillator shells (say $\eta_p$ and $\eta_n$ shells with $\eta_p \ne \eta_n$). In fact this agreement extends to all the nuclei considered in the Tables 1-14 (also to those considered in \cite{Ko-nhw}). Thus, the assumptions used by Bonatsos et al in \cite{Bona,Bona-4,Bona-5} imply good proxy-$SU(4)$ symmetry for nuclei with valence protons and neutrons occupying the same oscillator shell. This may give a justification for arguing that an effective $SU(4)$ symmetry is present in heavy nuclei as 
suggested in \cite{su4-1,su4-2}; see also \cite{su4-3,su4-4}. This observation certainly needs further investigations as it is important to identify the nature of this effective $SU(4)$ symmetry in heavy nuclei.

\item In Figs. 1 and 2 shown are the results for $\beta$ vs N for Te to
Pt nuclei and also for all the N=Z isotopes of Te to Pt. It is clearly seen that for all nuclei the hw and nhw-ee irreps give essentially the
same value for the $\beta$ parameter. Therefore, $\beta$ value will not be sensitive to the mixing of hw and nhw-ee irreps. In general, mixing of low-lying $SU(3)$ irreps will not affect $\beta$ values. Also as seen from the Figs. 1 and 2, the variation of $\beta$ is smooth and essentially parabolic
with maximum for N=64 (mid-shell region). All other variations seen with respect to change in Z value are as expected with Te and Xe nuclei having few valence protons and similarly, Os and Pt nuclei having few proton holes. However, as seen from Figs. 3 and 4, the asymmetry parameter $\gamma$ shows clear differences
in its value for hw and nhw-ee irreps. Also, there are clear fluctuations in $\gamma$ vs N. It is also seen that the mixing of hw and nhw-ee irreps affect significantly $\gamma$ values. Experimental tests of the results in Figs. 1-4 will be valuable.

\item In general to study triaxial shapes in nuclei it is necessary to mix the hw and nhw-ee irreps (may be also hw and all nhw irreps). A simple assumption of 50\% mixing is used in \cite{Bona,Bona-4,Bona-5}. Same assumption is used in Figs. 3 and 4. However, a better option is to use the quadratic Casimir operator of $SO(15)$ in $U(15) \supset SO(15)$ and this is same as the isoscalar plus isovector pairing operator in the proxy $\eta=4$ shell \cite{Bahri,KoCas,Kali}). This operator will mix $SU(3)$ irreps but keeps $SU_{ST}(4)$ to be a good symmetry. 

\item Besides the $SO(15)$ subalgebra, $U(15)$ also admits $U(5)$ and $U(6)$ limits with
$$
U(15) \supset U(5) \supset SO(5) \supset SO(3)\;,\;\;\;
U(15) \supset U(6) \supset Sp(6) \supset SO(3)\;.
$$
The generators of the various algebras in the $U(5)$ and $U(6)$ chains can be written down using the results derived in the context of $sdg$IBM \cite{JMP}. Similarly, it is also
possible to construct the quadratic Casimir invariants of the $U(5)$, $SO(5)$, $U(6)$ and $Sp(6)$ algebras above; see \cite{JMP}. These also will mix the $SU(3)$ irreps but keep $SU_{ST}(4)$ to be a good symmetry. Calculations involving these operators will be interesting and may provide new insights into proxy-$SU(4)$ symmetry. For completeness, we mention that $U(15)$ also admits $U(15) \supset U(14) \supset SO(14) \supset \ldots$ and $U(15) \supset \l[U(6) \oplus U(9)\r] \supset \ldots$ and 
$U(15) \supset \l[U(5) \oplus U(10)\r] \supset \ldots$ algebras \cite{JMP}.

\item Experimental evidence for the goodness of proxy-$SU(4)$ symmetry for nuclei with $52 \le \mbox{Z,N} \le 78$ is expected 
to come from binding energy systematics and GT strength distributions. Several of the nuclei listed in Tables 1-14 are the subject of some recent experiments and also theoretical studies using the relativistic mean-field theory and for cluster radioactivity; see for example Refs. \cite{Xe108,Te104,XeGd,Xe-rmf,Ba112}. Nuclei in Tables 1 to 14 consist of a wide variety with neutron rich nuclei, neutron deficient nuclei, N=Z nuclei, N and/or Z close to closed shells and so on. Let us add that it is important to analyze available experimental data and also examine the results of various theoretical models for nuclei with $52 \le \mbox{Z,N} \le 78$, both with the aim of finding signatures of proxy-$SU(4)$ symmetry in these nuclei.

\item There is now renewed interest in Wigner's $SU_{ST}(4)$ symmetry \cite{jpd-su4a,jpd-su4b}. Therefore, besides proxy-$SU(4)$ symmetry, it will be interesting to apply the pseudo-$SU(4)$ symmetry \cite{psu4-1,psu4-2,psu4-3} and also the IBM-4 model with $SU_{ST}(4)$ symmetry \cite{ibm-su4} for the nuclei with $52 \le \mbox{Z,N} \le 78$ and make a comparitive study.
 
\end{enumerate}

\newpage
\captionsetup{justification=Centering}
\begin{table}[h!]
\centering
\caption{{\footnotesize{Low-lying $SU(3)$ irreps (giving hw irrep, nhw irrep, nnhw irrep etc.) $(\la,\mu)$ for $^{104-130}$Te even-even isotopes. In the table results are shown as $\Gamma(\la,\mu)$ where $\Gamma$ ($\ge 1$) is the multiplicity of the $SU(3)$ irrep $(\la,\mu)$. Listed only are those $(\la,\mu)$ irreps with $2\la+\mu=\epsilon_H$, $\epsilon_H-3$ and $\epsilon_H-6$. Shown in the table are also number of valence nucleons $m$, isospin $T$ and the $U(15)$ irrep $\l\{f\r\}$. See text for further details.}}}
{\scriptsize{
\begin{tabular}{l}
\hline
$^{104}$Te $\;:\;\;(m=4,T=0)\;:\;\;\l\{f\r\}=\l\{4\r\}$ \\
\verb|  1(16  0)   1(12  2) | \\
$^{106}$Te $\;:\;\;(m=6,T=1)\;:\;\;\l\{f\r\}=\l\{4,2\r\}$ \\
\verb|  1(20  2)   1(18  3)   1(19  1)   3(16  4)   2(17  2)   2(18  0) | \\
$^{108}$Te $\;:\;\;(m=8,T=2)\;:\;\;\l\{f\r\}=\l\{4,2^2\r\}$ \\     
\verb|  1(26  0)   1(23  3)   1(24  1)   1(20  6)   2(21  4)   5(22  2) | \\
$^{110}$Te $\;:\;\;(m=10,T=3)\;:\;\;\l\{f\r\}=\l\{4,2^3\r\}$ \\     
\verb|  1(26  4)   1(28  0)   2(24  5)   3(25  3)   2(26  1)   1(21  8)   7(22  6) | \\
\verb|  8(23  4)  12(24  2)   2(25  0) | \\
$^{112}$Te $\;:\;\;(m=12,T=4)\;:\;\;\l\{f\r\}=\l\{4,2^4\r\}$ \\       
\verb|  1(28  4)   1(30  0)   1(25  7)   2(26  5)   3(27  3)   3(28  1)   1(22 10) | \\ 
\verb|  3(23  8)  10(24  6)  11(25  4)  15(26  2)   3(27  0) | \\
$^{114}$Te $\;:\;\;(m=14,T=5)\;:\;\;\l\{f\r\}=\l\{4,2^5\r\}$ \\         
\verb|  1(32  0)   1(28  5)   1(29  3)   1(30  1)   1(24 10)   1(25  8)   5(26  6) | \\
\verb|  5(27  4)   8(28  2) | \\
$^{116}$Te $\;:\;\;(m=16,T=6)\;:\;\;\l\{f\r\}=\l\{4,2^6\r\}$ \\           
\verb|  1(28  6)   1(30  2)   1(25  9)   3(26  7)   4(27  5)   4(28  3)   3(29  1) | \\
\verb|  2(22 12)   4(23 10)  14(24  8)  17(25  6)  24(26  4)  15(27  2)  10(28  0) | \\
$^{118}$Te $\;:\;\;(m=18,T=7)\;:\;\;\l\{f\r\}=\l\{4,2^7\r\}$ \\         
\verb|  1(26  8)   2(28  4)   1(30  0)   1(23 11)   3(24  9)   5(25  7)   7(26  5) | \\
\verb|  7(27  3)   4(28  1)   2(20 14)   4(21 12)  14(22 10)  20(23  8)  35(24  6) | \\
\verb| 30(25  4)  32(26  2)   6(27  0) | \\
$^{120}$Te $\;:\;\;(m=20,T=8)\;:\;\;\l\{f\r\}=\l\{4,2^8\r\}$ \\           
\verb|  1(26  6)   1(28  2)   1(22 11)   2(23  9)   3(24  7)   5(25  5)   5(26  3) | \\
\verb|  3(27  1)   1(18 16)   1(19 14)   6(20 12)   9(21 10)  19(22  8)  21(23  6) | \\
\verb| 29(24  4)  17(25  2)  11(26  0) | \\
$^{122}$Te $\;:\;\;(m=22,T=9)\;:\;\;\l\{f\r\}=\l\{4,2^9\r\}$ \\             
\verb|  1(28  0)   1(23  7)   1(24  5)   1(25  3)   1(26  1)   1(18 14)   1(19 12) | \\
\verb|  4(20 10)   4(21  8)   9(22  6)   6(23  4)   9(24  2) | \\
$^{124}$Te $\;:\;\;(m=24,T=10)\;:\;\;\l\{f\r\}=\l\{4,2^{10}\r\}$ \\               
\verb|  1(20  8)   1(22  4)   1(24  0)   1(16 13)   2(17 11)   3(18  9)   5(19  7) | \\
\verb|  5(20  5)   4(21  3)   3(22  1)   1(12 18)   1(13 16)   5(14 14)   8(15 12) | \\
\verb| 17(16 10)  19(17  8)  28(18  6)  20(19  4)  21(20  2)   3(21  0) | \\
$^{126}$Te $\;:\;\;(m=26,T=11)\;:\;\;\l\{f\r\}=\l\{4,2^{11}\r\}$ \\                 
\verb|  1(14 12)   2(16  8)   1(17  6)   2(18  4)   2(20  0)   1(11 15)   2(12 13) | \\
\verb|  4(13 11)   6(14  9)   8(15  7)   8(16  5)   7(17  3)   4(18  1)   1( 8 18) | \\
\verb|  1( 9 16)   7(10 14)  10(11 12)  21(12 10)  23(13  8)  34(14  6)  24(15  4) | \\
\verb| 25(16  2)   3(17  0) | \\
$^{128}$Te $\;:\;\;(m=28,T=12)\;:\;\;\l\{f\r\}=\l\{4,2^{12}\r\}$ \\                   
\verb|  1(10 12)   2(12  8)   1(13  6)   2(14  4)   2(16  0)   1( 8 13)   2( 9 11) | \\
\verb|  3(10  9)   5(11  7)   5(12  5)   4(13  3)   3(14  1)   2( 6 14)   3( 7 12) | \\
\verb|  9( 8 10)  10( 9  8)  17(10  6)  12(11  4)  14(12  2)   1(13  0) | \\
$^{130}$Te $\;:\;\;(m=30,T=13)\;:\;\;\l\{f\r\}=\l\{4,2^{13}\r\}$ \\                     
\verb|  1( 8  8)   1(10  4)   1(12  0)   1( 7  7)   1( 8  5)   1( 9  3)   1(10  1) | \\
\verb|  1( 4 10)   1( 5  8)   4( 6  6)   2( 7  4)   4( 8  2) | \\
\hline
\end{tabular}
}}
\end{table}
\newpage
\begin{table}[h!]
\centering
\caption{{\footnotesize{Low-lying $SU(3)$ irreps (giving hw irrep, nhw irrep, nnhw irrep etc.) $(\la,\mu)$ for $^{106-132}$Xe even-even isotopes. In the table results are shown as $\Gamma(\la,\mu)$ where $\Gamma$ ($\ge 1$) is the multiplicity of the $SU(3)$ irrep $(\la,\mu)$. Listed only are those $(\la,\mu)$ irreps with $2\la+\mu=\epsilon_H$, $\epsilon_H-3$ and $\epsilon_H-6$. Shown in the table are also number of valence nucleons $m$, isospin $T$ and the $U(15)$ irrep $\l\{f\r\}$. In addition $\l\{f_h\r\}$ is also given when appropriate. See text for further details.}}}
{\scriptsize{

}}
\end{table}
\newpage
\begin{table}[h!]
\centering
\caption{{\footnotesize{Low-lying $SU(3)$ irreps (giving hw irrep, nhw irrep, nnhw irrep etc.) $(\la,\mu)$ for $^{108-134}$Ba even-even isotopes. See caption to Table 2 for further details.}}}
{\scriptsize{
%
}}
\end{table}
\newpage
\begin{table}[h!]
\centering
\caption{{\footnotesize{Low-lying $SU(3)$ irreps (giving hw irrep, nhw irrep, nnhw irrep etc.) $(\la,\mu)$ for $^{110-136}$Ce even-even isotopes. See caption to Table 2 for further details.}}}
{\scriptsize{
%
}}
\end{table}
\newpage
\begin{table}[h!]
\centering
\caption{{\footnotesize{Low-lying $SU(3)$ irreps (giving hw irrep, nhw irrep, nnhw irrep etc.) $(\la,\mu)$ for $^{112-138}$Nd even-even isotopes. See caption to Table 2 for further details.}}}
{\scriptsize{
%
}}
\end{table}
\newpage
\begin{table}[h!]
\centering
\caption{{\footnotesize{Low-lying $SU(3)$ irreps (giving hw irrep, nhw irrep, nnhw irrep etc.) $(\la,\mu)$ for $^{114-140}$Sm even-even isotopes. See caption to Table 2 for further details.}}}
{\scriptsize{
%
}}
\end{table}
\newpage
\begin{table}[h!]
\centering
\caption{{\footnotesize{Low-lying $SU(3)$ irreps (giving hw irrep, nhw irrep, nnhw irrep etc.) $(\la,\mu)$ for $^{116-142}$Gd even-even isotopes. See caption to Table 2 for further details.}}}
{\scriptsize{
%
}}
\end{table}
\newpage
\begin{table}[h!]
\centering
\caption{{\footnotesize{Low-lying $SU(3)$ irreps (giving hw irrep, nhw irrep, nnhw irrep etc.) $(\la,\mu)$ for $^{118-144}$Dy even-even isotopes. See Caption to Table 2 for further details.}}}
{\scriptsize{
%
}}
\end{table}
\newpage
\begin{table}[h!]
\centering
\caption{{\footnotesize{Low-lying $SU(3)$ irreps (giving hw irrep, nhw irrep, nnhw irrep etc.) $(\la,\mu)$ for $^{120-146}$Er even-even isotopes. See Caption to Table 2 for further details.}}}
{\tiny{
%
}}
\end{table}
\newpage
\begin{table}[h!]
\centering
\caption{{\footnotesize{Low-lying $SU(3)$ irreps (giving hw irrep, nhw irrep, nnhw irrep etc.) $(\la,\mu)$ for $^{122-148}$Yb even-even isotopes. See caption to Table 2 for further details.}}}
{\scriptsize{
%
}}
\end{table}
\newpage
\begin{table}[h!]
\centering
\caption{{\footnotesize{Low-lying $SU(3)$ irreps (giving hw irrep, nhw irrep, nnhw irrep etc.) $(\la,\mu)$ for $^{124-150}$Hf even-even isotopes. See Caption to Table 2 for further details.}}}
{\scriptsize{
%
}}
\end{table}
\newpage
\begin{table}[h!]
\centering
\caption{{\footnotesize{Low-lying $SU(3)$ irreps (giving hw irrep, nhw irrep, nnhw irrep etc.) $(\la,\mu)$ for $^{126-152}$W even-even isotopes. See Caption to Table 2 for further details.}}}
{\scriptsize{
%
}}
\end{table}
\newpage
\begin{table}[h!]
\centering
\caption{{\footnotesize{Low-lying $SU(3)$ irreps (giving hw irrep, nhw irrep, nnhw irrep etc.) $(\la,\mu)$ for $^{128-154}$Os even-even isotopes. See caption to Table 2 for further details.}}}
{\scriptsize{
%
}}
\end{table}
\newpage
\begin{table}[h!]
\centering
\caption{{\footnotesize{Low-lying $SU(3)$ irreps (giving hw irrep, nhw irrep, nnhw irrep etc.) $(\la,\mu)$ for $^{130-156}$Pt even-even isotopes. See caption to Table 2 for further details.}}}
{\tiny{
%
}}
\end{table}
\newpage

\begin{figure}
\includegraphics[height=20cm]{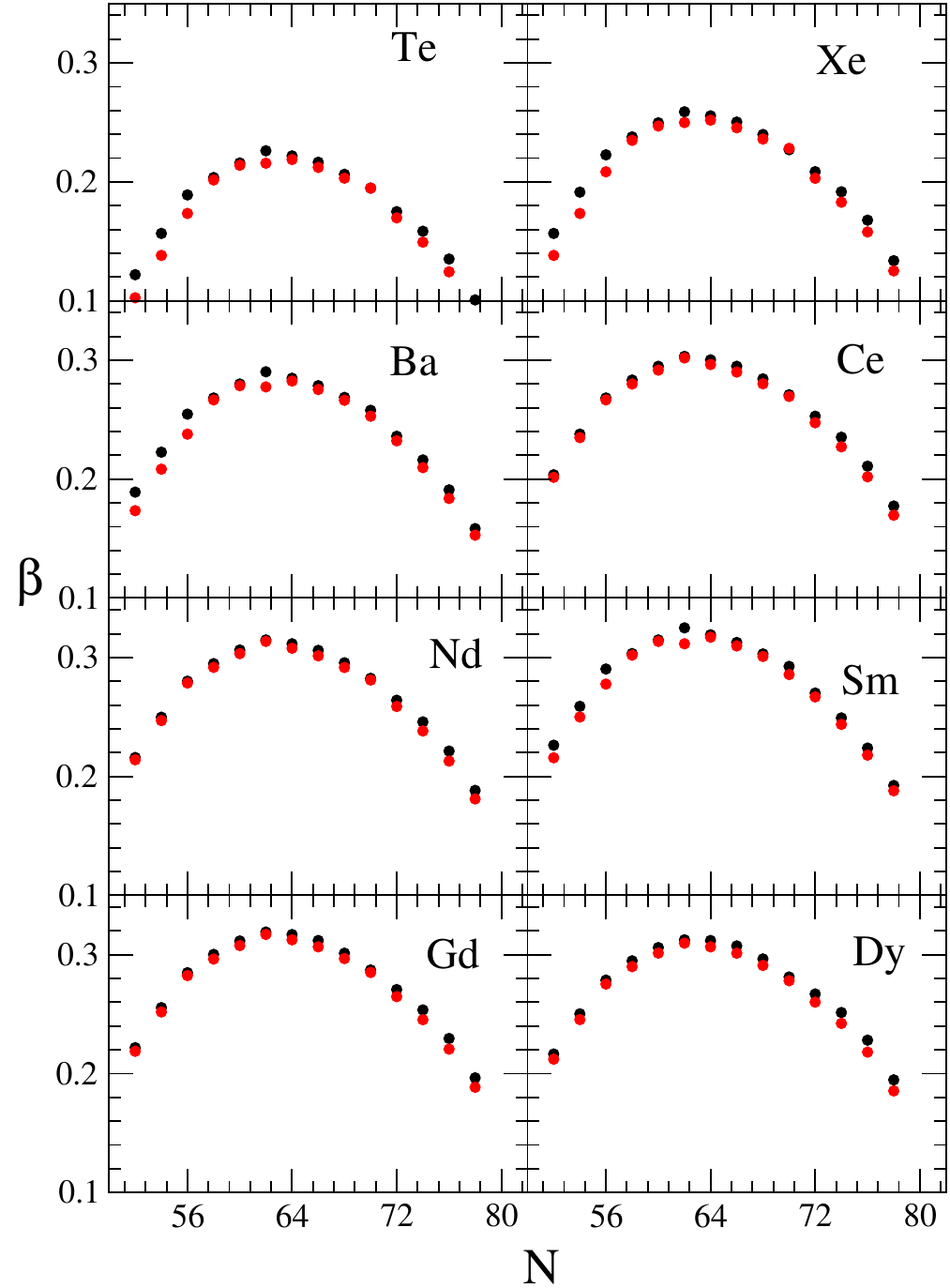}
	\caption{Deformation parameter $\beta$ vs N (neutron number)
for Te, Xe, Ba, Ce, Nd, Sm, Gd and Dy nuclei. The filled black circles are results for hw $SU(3)$ irrep and red filled circles are results for nhw-ee irrep. Values of the deformation parameter $\beta$ are almost same for hw and nhw-ee irrep. Thus, mixing of these irreps will not change much the $\beta$ value. See text for further details.}
\label{bet1}
\end{figure}

\newpage
\begin{figure}
\includegraphics[height=20cm]{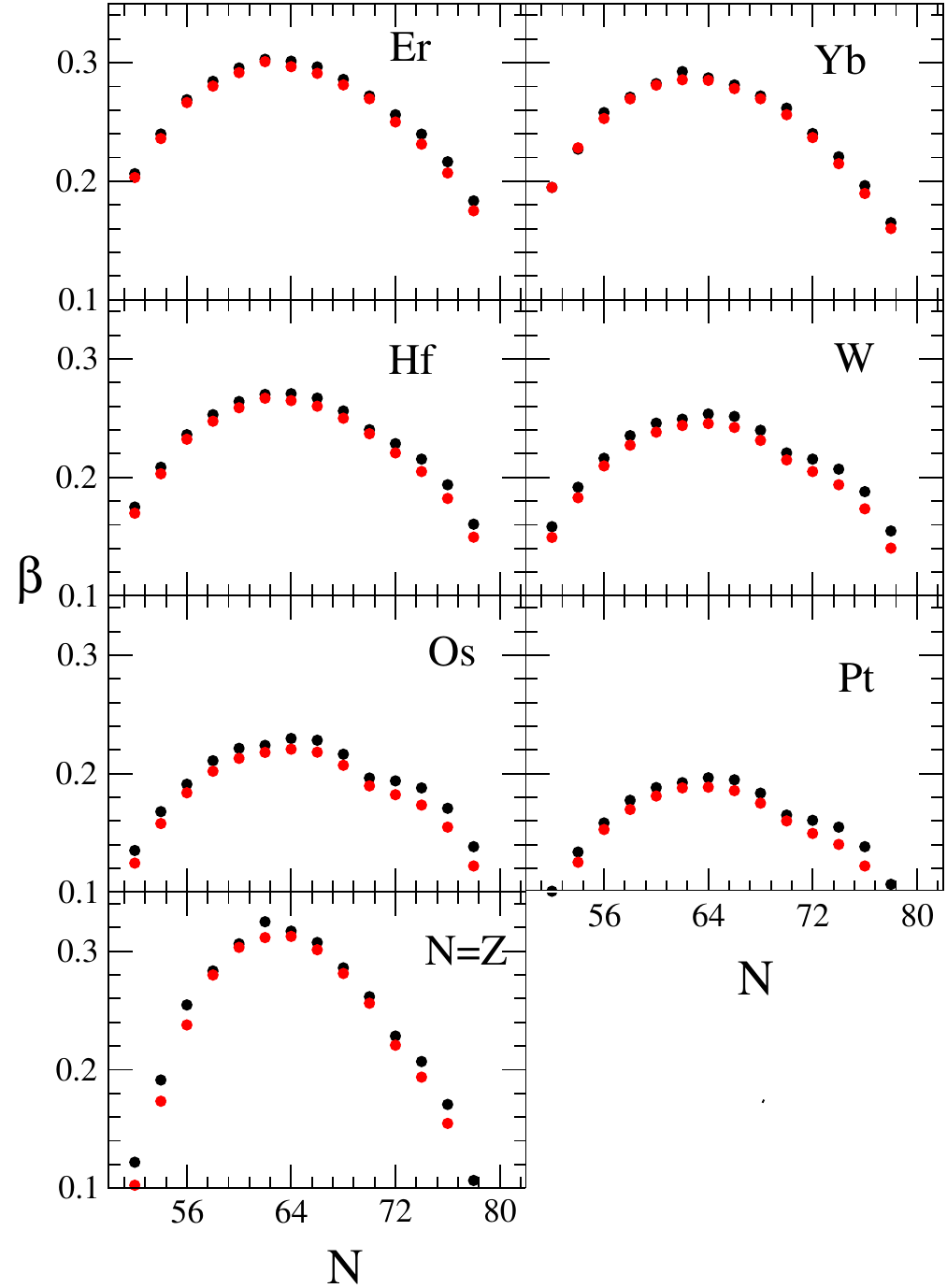}
	\caption{Deformation parameter $\beta$ vs N (neutron number)
for Er, Yb, Hf, W, Os and Pt nuclei. Shown also are the results for
N=Z nuclei with $52 \le \mbox{Z} \le 78$.  The filled black circles are results for hw $SU(3)$ irrep and red filled circles are results for nhw-ee irrep. Values of the deformation parameter $\beta$ are almost same for hw and nhw-ee irrep. Thus, mixing of these irreps will not change much the $\beta$ value. See text for further details.}
\label{bet2}
\end{figure}

\newpage

\begin{figure}
\includegraphics[height=20cm]{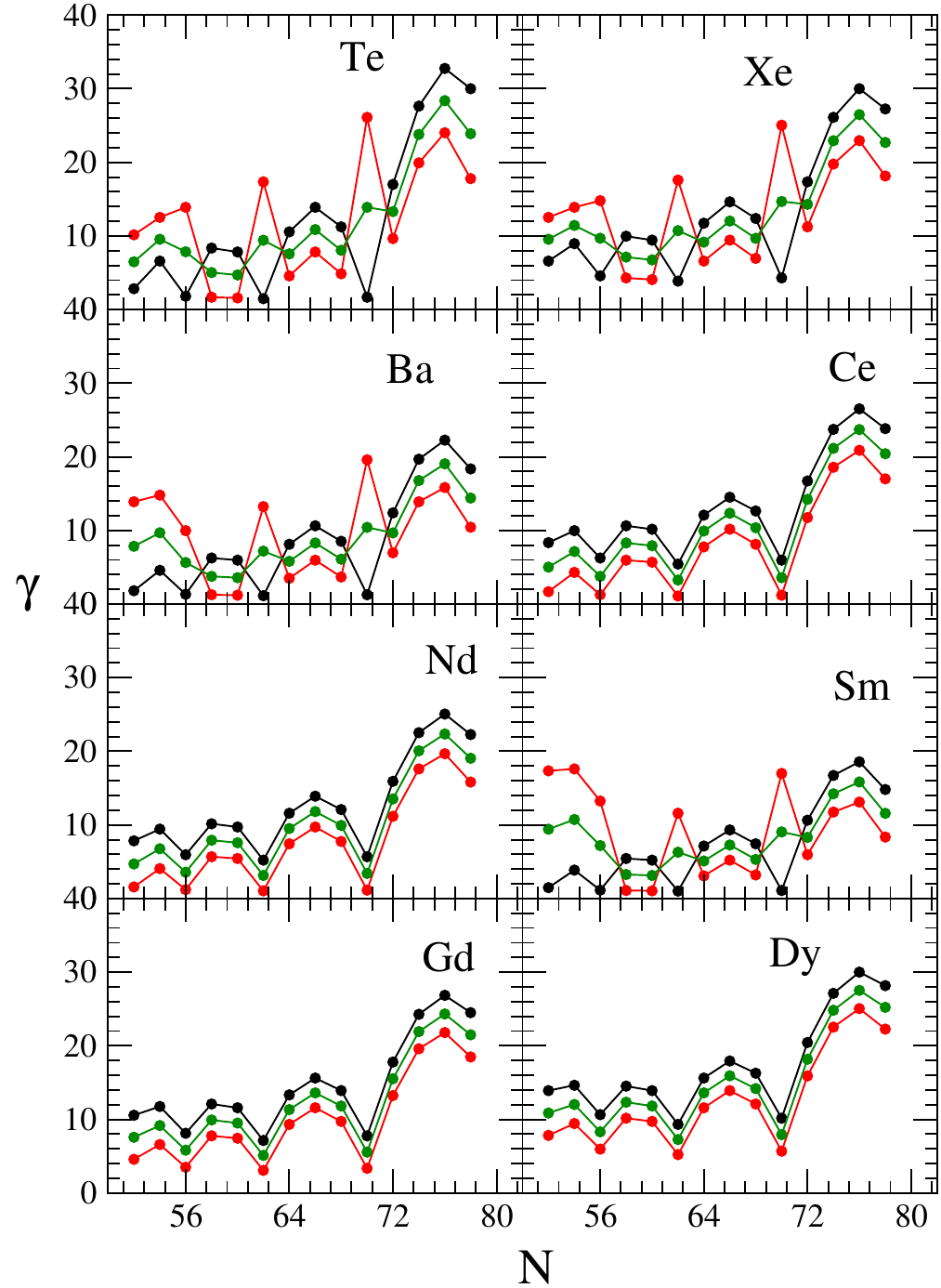}
	\caption{Deformation parameter $\gamma$ vs N (neutron number)
for Te, Xe, Ba, Ce, Nd, Sm, Gd and Dy nuclei. The filled black circles are results for hw $SU(3)$ irrep, red filled circles are results for nhw-ee irrep and green filled circles are with mixing (50\% each) of the hw and nhw-ee irreps. Values of the defomation parameter $\gamma$ are quite different for hw and nhw-ee irrep. Thus, as seen from the figure, mixing of these irreps will change the $\gamma$ value. See text for further details.}
\label{gam1}
\end{figure}

\newpage
\begin{figure}
\includegraphics[height=20cm]{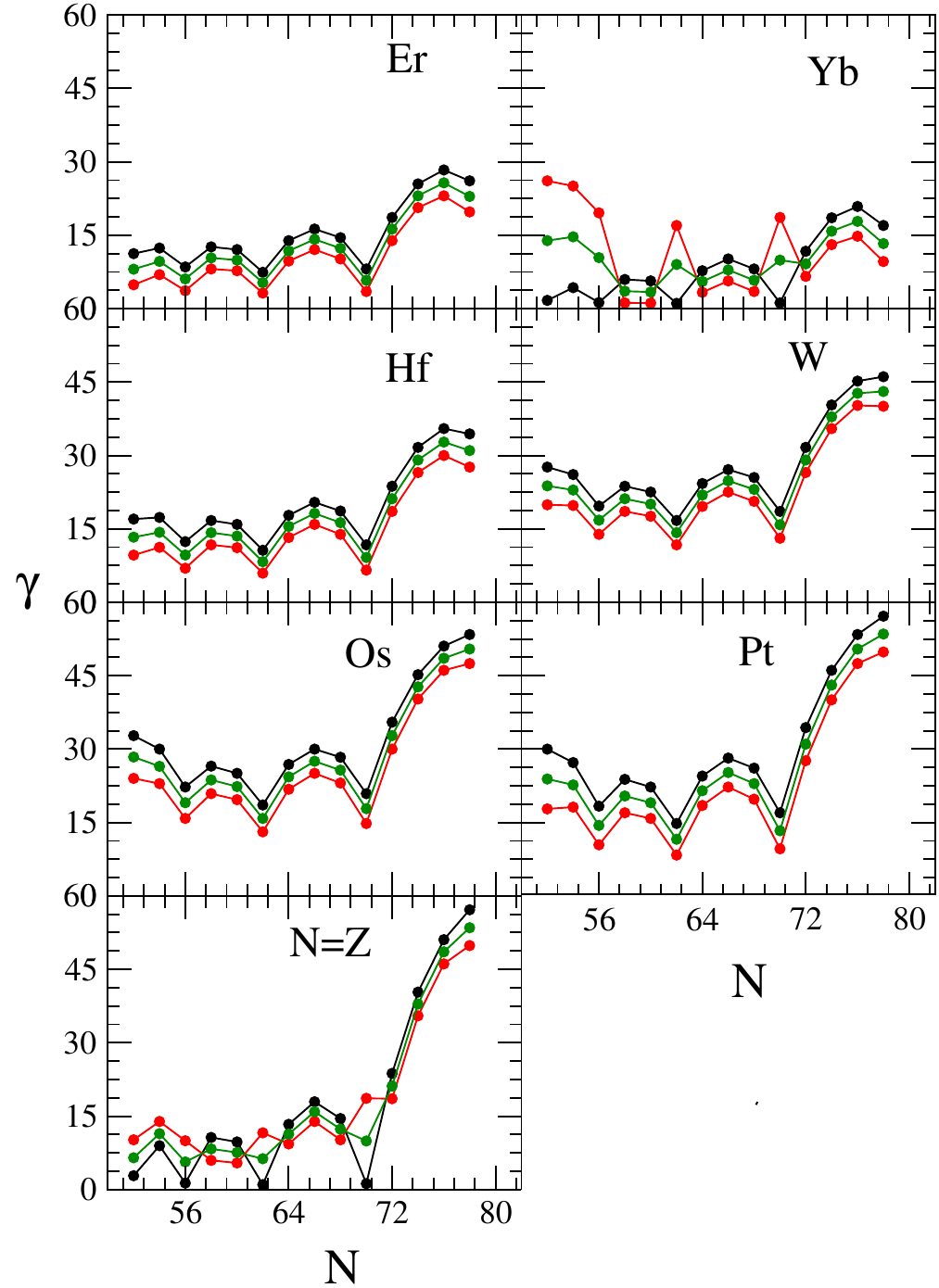}
	\caption{Deformation parameter $\gamma$ vs N (neutron number)
for Er, Yb, Hf, W, Os and Pt nuclei. Shown also are the results for
N=Z nuclei with $52 \le \mbox{Z} \le 78$. The filled black circles are results for hw $SU(3)$ irrep, red filled circles are results for nhw-ee irrep and green filled circles are with mixing (50\% each) of the hw and nhw-ee irreps. Values of the defomation parameter $\gamma$ are quite different for hw and nhw-ee irrep. Thus, as seen from the figure, mixing of these irreps will change the $\gamma$ value. See text for further details.}
\label{gam2}
\end{figure}

\newpage

\section{Reductions for boson systems}

The procedure used for shell model $SU(3)$ limits I, II and III, as described in Section 2, also applies easily to the $SU(3)$ limit of various interacting boson models (IBM's). For bosons with no internal degrees of freedom we have for example $sd$IBM with $U(6) \supset SU(3)$ \cite{Iac}, $sdpf$IBM with $U(16) \supset [U(6) \supset SU(3)] \oplus [U(10) \supset SU(3)]$ \cite{sdpf} and $sdg$IBM with $U(15) \supset SU(3)$ \cite{sdg}.
Thus, in general just as III, we have $U(\can) \supset SU(3)$ with
$\can=(\eta+1)(\eta+2)/2$ and $\eta=2,3,4$ being important (in some applications $\eta=5$ and $6$ are also useful \cite{KMC}). The only difference is that here we need to consider totally symmetric irreps $\{m\}$ of $U(\can)$ for an $m$ boson systems.
Then, we can put maximum of $m$ number of bosons in any $(n_z,n_x,n_y)$ sp state. Now, repeating the same procedure as in 
Section 2.2 we will obtain $\{m\}_{U(\can)}$ reductions to $SU(3)$ irreps,
\be
\l\{m\r\}_{U(\can)} = \dis\sum_{\la_s,\mu_s} \Gamma_{(\la_s,\mu_s)} \; (\la_s,\mu_s)\;.
\label{eq22}
\ee
Note that $(\la_s,\mu_s)$ are all the irreps for $m$ bosons and
$\la_s=f_1-f_2$, $\mu_s=f_2-f_3$ with $f_1+f_2+f_3=m\eta$ and
$f_1 \ge f_2 \ge f_3 \ge 0$. In stead of the procedure used in Section 2.2, there are many other methods for obtaining the $SU(3)$
reductions of $\{m\}$ of $U(\can)$; see \cite{su3-book}. We have generated tabulations for $\eta=3$ with $m=1-50$, $\eta=4$ with $m=1-48$, $\eta=5$ with $m=1-26$ and $\eta=6$ with $m=1-19$. Note that for $\eta=2$ a simple analytical solution is available giving,
\be
\barr{l}
\l\{m\r\}_{U(6)} = \dis\sum_{\la_s,\mu_s}  (\la_s,\mu_s)\;; \\
\la_s=2(f_1-f_2),\;\;\;\mu_s=2(f_2-f_3),\;\;\;f_1+f_2+f_3=m,\;\;\;f_2 \ge f_2 \ge f_3 \ge 0 \;.
\earr \label{eq23}
\ee
Going further, as needed for IBM-2 models with $F$-spin \cite{Iac}, it is easy to obtain the $SU(3)$ irreps. Just as
II in Section 2, here the group chain is \cite{Iac}
$$
U(2\can) \supset \l[U(\can) \supset SU(3)\r] \otimes SU_F(2)
$$
with the $U(\can)$ irreps being two rowed $\{f_1,f_2\}$ with
$f_1+f_2=m$ and $f_1-f_2=2F$. Now, the reduction of any given
$\{f_1,f_2\}$ (i.e. given $(m,F)$) to $SU(3)$ irreps follows by expanding $\{f_1,f_2\}$ to symmetric irreps using
\be
\l\{f_1,f_2\r\} = \l\{f_1\r\}\times \l\{f_2\r\} - \l\{f_1+1\r\}\times \l\{f_2-1\r\}\;.
\label{eq24}
\ee
Now, using the reductions of symmetric irreps $\{m\}$ and Eq. (\ref{eq24}) together with the procedure used in Section 2.3
with give the final results. Tables for $\can=6$, 10 and 15, for some reasonable values of $m$ are easily generated and these results will be available upon request.

Going further, for IBM-3 models with isospin ($T$) degree of freedom \cite{KS,Van}, we have 
$$
U(3\can) \supset \l[U(\can) \supset SU(3)\r] \otimes \l[SU_T(3)
\supset SO_T(3)\r]
$$
Here, the $U(\can)$ irreps will be a three rowed partition $\{f_1,f_2,f_3\}$ with $f_1+f_2+f_3=m$ and $f_1 \ge f_2 \ge f_3 \ge 0$. For a
given $\{f_1,f_2,f_3\}$, the $SU_T(3)$ irrep is uniquely $(\la_T,\mu_T)$ with $\la_T=f_1-f_2$ and $\mu_T=f_2-f_3$. Also, reduction of $(\la_T,\mu_T)$ to $T$ follows simply from Eq. (\ref{eq20}) by replacing $L$ with $T$. Following Section 2.3.2 but using the symmetric irreps via the following $3 \times 3$ matrix
\be
\barr{l}
\l\{f_1,f_2,f_3\r\} = \l|\barr{ccc} \l\{f_1\r\} &  
\l\{f_1+1\r\} &  \l\{f_1+2\r\} \\
\l\{f_2-1\r\} &  \l\{f_2\r\} &  \l\{f_2+1\r\} \\ 
\l\{f_3-2\r\} &  \l\{f_3-1\r\} &  \l\{f_3\r\} \earr \r|\;,
\earr \label{eq25}
\ee
we will obtain the reduction of any $\{f_1,f_2,f_3\}_{U(\can)}$ to $SU(3)$ irreps. With $\can=6$, 10 and 15 it is possible to generate
tables for $U(\can) \supset SU(3)$ reductions for reasonable values of $m$
and all allowed $\{f_1,f_2,f_3\}$. This can be continued also for IBM-4
models \cite{ibm-su4,Iac,KS} with the group chain being
$$
U(6\can) \supset \l[U(\can) \supset SU(3) \supset SO(3)\r] \otimes \l[
U(6) \supset SO(6) \supset SO_S(3) \oplus SO_T(3)\r]\;.
$$
Note that the SO(6) chain here is equivalent to Wigner $SU(4)$ chain
$U(6) \supset SU(4) \supset SU_S(2) \oplus SU_T(2)$. Thus, in IBM-4
we need to deal with $U(\can)$ irreps that are six rowed. Therefore,
to obtain $SU(3)$ reductions Eq. (\ref{eq25}) will change to an equation with a $6 \times 6$ matrix.
 
\section{Conclusions}

Applying some simple rules, established in group theory literature for $SU(3)$ algebra, developed is a FORTRAN code that can be easily modified as required for obtaining $SU(3)$ irreps for a given number $m$ of nucleons in a given oscillator shell $\eta$. The method used is described in
Section 2 in some detail. Tabulations are generated (available on request)for the reduction of one, two and four column Young tableaux of $U(\can)$ to $SU(3)$ irreps [$\can = (eta +1)(\eta +2)/2$] as described in Section 2. Results are
available for any $m$ up to $\eta=7$ and for $m \le 10$ for $\eta=8$. As an application of these tables, results for the reductions are presented, with $SU(3)$ irreps that correspond to $\epsilon_H$, $\epsilon_H-3$ and $\epsilon_H-6$, in Section 3 in Tables 1-14 for $52 \le \mbox{Z,N} \le 78$ nuclei with proxy-$SU(4)$ symmetry. Then, we have nucleons in $\eta=4$ shell giving $U(15) \supset SU(3)$ and the
$U(15)$ irreps are four columned. Further, using the hw and nhw-ee $SU(3)$ irreps, calculated are the quadrupole deformation parameters $(\beta,\gamma)$ and Figs. 1-4 show the results. Several important questions that arise from the results in Tables 1-14 and Figs. 1-4 are
also discussed in Section 3. These include: (i) constructing operators
that distinguish $SU(3)$ irreps with same $(\la , \mu)$; (ii)  calculations using pairing operator that mixes hw and nhw irreps; (iii)
understanding the nature of a possible effective $SU(4)$ symmetry in heavy nuclei; (iv) analyzing available experimental data and also the results from other theoretical studies. Answering these needs further work in future. Finally, extension of the method used in Section 2 to various interacting boson
models of nuclei are briefly discussed in Section 4. We hope that the
results presented in this work will prompt further work on nuclei with $52 \le \mbox{Z,N} \le 78$. These may establish the goodness of proxy-$SU(4)$ symmetry in these nuclei and also understanding better the existence of an effective $SU(4)$ symmetry in heavy nuclei with valence protons and neutrons in different oscillator shells.
 
\ack

Thanks are due to D. Bonatsos for some useful correspondence. Thanks are also to R. Sahu and N.D. Chavda for help in preparing the manuscript. 


\end{document}